\documentclass[prd,twocolumn,showpacs]{revtex4}
\usepackage{graphicx}
\usepackage{bm}
\usepackage{dcolumn}
\usepackage{amsmath}
\usepackage{amssymb}
\usepackage{color}
\usepackage{url}

\usepackage{aas_macros}

\usepackage{hyperref}

\usepackage{subfigure}

\usepackage{color}
\newcommand{\red}{}

\def\aj{AJ}
\def\nat{Nature}
\def\araa{ARAA}
\def\prd{Phys Rev D}
\def\apj{ApJ}
\def\apjl{ApJL}
\def\apjs{ApJS}
\def\mnras{MNRAS}

\def\aap{A \& A}
\def\arcsec{{\rm Arcsec}}
\def\arcsech{{\rm Arcsech}}

\def\rh{R_{\rm h}}

\def\Mh{M_{\rm h}}
\def\sigmalos{\sigma_{\rm los}}
\def\vlos{v_{\rm los}}

\def\rs{r_{\rm s}}

\def\kms{\,{\rm km\,s^{-1}}}

\def\percent{\text{ per cent}}

\def\rhoDM{{\rho_{\rm DM}}}
\def\angle{{\theta}}
\def\amax{{\angle_{\rm max}}}
\def\acrit{{\angle_{\rm crit}}}

\def\percent{\text{ per cent}}

\newcommand{\Df}{{\rm D}}
\newcommand{\Jf}{{\rm J}}

\begin{document}

% Title & Authors
\title{Simple J-Factors and D-Factors for Indirect Dark Matter Detection}

\author{N.~W. Evans}
\email{nwe@ast.cam.ac.uk}
\affiliation{Institute of Astronomy, Madingley Rd, Cambridge, CB3 0HA}

\author{J.~L. Sanders}
\email{jls@ast.cam.ac.uk}
\affiliation{Institute of Astronomy, Madingley Rd, Cambridge, CB3 0HA}

\author{Alex Geringer-Sameth}
\email{alexgs@cmu.edu}
\affiliation{McWilliams Center for Cosmology, Department of Physics, Carnegie Mellon University, Pittsburgh, PA 15213}

\date{\today}

\begin{abstract}
{\red J-factors (or D-factors) describe the distribution of dark
  matter in an astrophysical system and determine the strength of the
  signal provided by annihilating (or decaying) dark matter
  respectively}. We provide simple analytic formulae to calculate the
J-factors for spherical cusps obeying the empirical relationship
between enclosed mass, velocity dispersion and half-light radius. We
extend the calculation to the spherical Navarro-Frenk-White (NFW)
model, and demonstrate that our new formulae give accurate results in
comparison to more elaborate Jeans models driven by Markov Chain Monte
Carlo methods. Of the known ultrafaint dwarf spheroidals, we show that
Ursa Major II, Reticulum II, Tucana II and Horologium I have the
largest J-factors and so provide the most promising candidates for
indirect dark matter detection experiments. Amongst the classical
dwarfs, Draco, Sculptor and Ursa Minor have the highest J-factors. We
show that the behaviour of the J-factor as a function of integration
angle can be inferred for general dark halo models with inner slope
$\gamma$ and outer slope $\beta$. The central and asymptotic behaviour
of the J-factor curves are derived as a function of the dark halo
properties. Finally, we show that models obeying the empirical
relation on enclosed mass and velocity dispersion have J-factors that
are most robust at the integration angle equal to the projected
half-light radius of the dSph divided by heliocentric distance. {\red
  For most of our results, we give the extension to the D-factor which
  is appropriate for the decaying dark matter picture.}
\end{abstract}


\maketitle

\section{Introduction}

The dwarf galaxies surrounding the Milky Way are the most extreme dark
matter dominated objects known to us with central mass to light ratios
typically of the order of tens to hundreds~\citep[e.g.,][]{Ma98,Kl02}.
Additionally, little or no emission has been detected in wavebands
other than the optical, and so the intrinsic astrophysical backgrounds
are low.  This makes the dSphs attractive targets to look for signals
of dark matter annihilation products~\citep{La90,Ev04,Cu06,Co07}.

The $\gamma$ ray differential flux from dark matter annihilation
measured within a solid angle $\Delta \Omega$ is~\citep[see
  e.g.,][]{Be98,Ev04}
\begin{equation}
{d \phi_\gamma \over d E_\gamma} = \phi^{\rm {PP}}(E) \times \Jf(\Delta
\Omega).
\end{equation}
The first term on the right-hand side depends on the particle physics
(i.e. dark matter particle mass, annihilation cross section, and
Standard Model final states).  The second term is the astrophysical
factor, or J-factor, and encapsulates the distribution of dark matter
within the system of interest:
\begin{equation}
\Jf = \int\int \rhoDM^2(\ell, \Omega) d\ell d\Omega.
\label{eq:Jfactor}
\end{equation}
It is therefore the square of the dark matter density integrated along
the line of sight and over the solid angle of the sky corresponding to the observation.

For the dSphs, the dark matter density is not known, but can be
constrained from the line of sight velocities of individual stars.
The spherical Jeans equations are often used to relate the velocities
of the stars to the underlying dark matter distribution. Nowadays,
this is often explored in a Bayesian framework using Monte Carlo
techniques, {\red so that the calculation of the J-factors requires
significant computational
effort~\citep{Ab10,Ch11,Wa11,Ge15b,Bo15a,Mar15}.}

However, it is reasonable to look for a simpler way of computing
J-factors for two reasons. First, for many of the recently discovered
ultrafaint dwarf galaxies, there are few stars with line of sight
velocities. In some cases, such as Ursa Major II, the giant branch
itself is so sparse that there are very few target stars for
spectroscopy~\citep{Zu06}.  Given such fundamental limitations on the
observational data, the extensive exploration of model-space in
conventional Jeans analyses may be needlessly elaborate.  Second, an
entirely characteristic feature of the dSphs is that they are
flattened.  In fact, some of the ultrafaints are very highly flattened
with ellipticities exceeding $0.5$, such as Ursa Major
II~\citep{Zu06}, Hercules~\citep{De12}, Ursa Major I~\citep{Ma08} and
Reticulum II~\citep{Ko15}. Therefore, the underlying physical model of
a spherical dark halo containing a round distribution of stars
satisfying the Jeans equation may fail to capture important aspects of
the physics.  This leaves the value of computationally intensive
approaches based on sphericity open to question, {\red as they may
  suffer from systematic effects when applied to flattened or triaxial
  systems.  While there have already been some investigations of
  axisymmetry and triaxiality~\citep{La13,Bo15a}, the implications of
  these effects on the dark matter annihilation signal warrants a
  systematic study, which we provide elsewhere~\citep{Sa16}.}

{\red The question as to whether dSphs have cored or cusped dark
  matter densities has been investigated intensively over the past few
  years.  The latest simulations have reached a resolution where the
  effect of individual supernovae can be modelled, making them much
  less sensitive to the details of the `sub-grid' star formation and
  feedback physics~\citep{On15,Re15}. They suggest that dark matter
  cores of size comparable to the half-light radius of the stellar
  component should be present in objects like Carina and Fornax that
  have formed stars for a Hubble time. However, dSphs with patchy star
  formation may be more cuspy, with the ultra-faints possibly
  retaining a pristine cusp of the form originally proposed by
  Navarro, Frenk \& White (NFW)~\cite{NFW}. This picture is broadly
  consistent with simple energy constraints~\cite{Pe12}, as well as
  tidal arguments suggesting that if the ultra-faints had dark matter
  cores, then they would not survive for long on their current
  orbits~\cite{Pe10}.  All this suggests that, even if the larger
  classical dSphs are cored as implied by the current observational
  data on general grounds~\citep{Ag12}, the smaller dSphs and the
  ultra-faints are probably cusped with dark matter densities behaving
  like $\rhoDM \sim r^{-1}$ at small radii.}

Here, we provide a set of simple formulae for J-factors in spherical
cusped geometries. Our approach is entirely elementary, but we show
that it leads to formulae that are very competitive with more
laborious approaches. Section II concentrates on the cosmologically
motivated $1/r$ cusp exemplified by the NFW model. We
provide an analytic formula for the J-factor of the NFW model, as well
as new estimates for the classical and ultrafaint dwarfs, including
the newly discovered Horologium I, Grus I, Hydra II, and Pisces II.
Section III extends the work into more general cusped and cored dark
haloes and explores the possibilty of inferring halo structure from
J-factor profiles. In particular, we show how the behaviour of the
J-factor at small and large integration angles is controlled by the
halo parameters. Finally, we investigate the existence of a sweet spot
{\red -- namely an integration angle at which the value of the
  J-factor is reasonably insensitive to the unknowable aspects of the
  underlying dark halo profile~\citep{Ch11,Wa11}.}

\begin{table*}
\caption{Annihilation and decay factors for dwarf spheroidals: we
  quote the predictions of the NFW formula from
  equations~\eqref{eq:Jfactorfull} and~\eqref{eq:Dfactorfull} using
  $r_s=5R_\mathrm{half}$ at two angles -- the angle between the centre
  of the dwarf and an estimate of the distance to the outermost member
  star $(\angle_\mathrm{max})$ and $\theta=0.5^\circ$. The dwarfs in
  the top section are classical dwarfs and those in the middle section
  are those ultrafaint dwarfs that have $J$ and $D$ estimates from the
  literature. The bottom section shows those ultrafaint dwarfs without
  pre-existing $J$ or $D$ estimates. For these we adopt
  $\angle_\mathrm{max}=0.5^\circ$.}  \begin{tabular}{llccccc}
\hline
\hline
Name & Distance & $\theta_\mathrm{max}$ & $\log_{10} J(\theta_\mathrm{max})$ & $\log_{10} J(0.5^\circ)$ & $\log_{10} D(\theta_\mathrm{max})$ & $\log_{10} D(0.5^\circ)$\\ 
&[$\mathrm{kpc}$]& [$^\circ$] & [$\mathrm{GeV^2\,cm}^{-5}$] & [$\mathrm{GeV^2\,cm}^{-5}$] & [$\mathrm{GeV\,cm}^{-2}$] & [$\mathrm{GeV\,cm}^{-2}$]\\
\hline
Carina&$105\pm6$ & $1.26$&$18.03_{-0.34}^{+0.34}$&$17.99_{-0.34}^{+0.34}$&$18.37_{-0.17}^{+0.17}$&$17.98_{-0.34}^{+0.34}$\\
Draco&$76\pm6$ & $1.3$&$18.92_{-0.25}^{+0.25}$&$18.86_{-0.24}^{+0.24}$&$18.82_{-0.12}^{+0.12}$&$18.39_{-0.25}^{+0.25}$\\
Fornax&$147\pm12$ & $2.61$&$18.27_{-0.17}^{+0.17}$&$18.15_{-0.16}^{+0.16}$&$19.04_{-0.09}^{+0.09}$&$18.26_{-0.17}^{+0.17}$\\
Leo I&$254\pm15$ & $0.45$&$17.80_{-0.28}^{+0.28}$&$17.80_{-0.28}^{+0.28}$&$17.84_{-0.14}^{+0.14}$&$17.89_{-0.28}^{+0.28}$\\
Leo II&$233\pm14$ & $0.23$&$17.41_{-0.25}^{+0.25}$&$17.44_{-0.25}^{+0.25}$&$17.31_{-0.12}^{+0.12}$&$17.62_{-0.25}^{+0.25}$\\
Sculptor&$86\pm6$ & $1.94$&$18.73_{-0.29}^{+0.29}$&$18.65_{-0.29}^{+0.29}$&$18.93_{-0.15}^{+0.15}$&$18.33_{-0.29}^{+0.29}$\\
Sextans&$86\pm4$ & $1.7$&$18.04_{-0.29}^{+0.29}$&$17.87_{-0.29}^{+0.29}$&$18.76_{-0.15}^{+0.15}$&$18.07_{-0.29}^{+0.29}$\\
Ursa Minor&$76\pm3$ & $1.37$&$19.18_{-0.24}^{+0.24}$&$19.15_{-0.24}^{+0.25}$&$18.84_{-0.12}^{+0.12}$&$18.45_{-0.24}^{+0.24}$\\
\hline
Bo\"otes I&$66\pm2$ & $0.47$&$16.64_{-0.38}^{+0.64}$&$16.65_{-0.38}^{+0.64}$&$17.25_{-0.19}^{+0.32}$&$17.28_{-0.38}^{+0.64}$\\
Coma Berenices&$44\pm4$ & $0.31$&$18.64_{-0.32}^{+0.32}$&$18.67_{-0.32}^{+0.33}$&$17.83_{-0.16}^{+0.16}$&$18.06_{-0.32}^{+0.32}$\\
Canes Venatici I&$218\pm10$ & $0.53$&$17.27_{-0.11}^{+0.11}$&$17.27_{-0.11}^{+0.11}$&$17.81_{-0.05}^{+0.05}$&$17.78_{-0.11}^{+0.11}$\\
Canes Venatici II&$160\pm4$ & $0.13$&$17.63_{-0.40}^{+0.40}$&$17.65_{-0.40}^{+0.40}$&$16.91_{-0.20}^{+0.20}$&$17.37_{-0.40}^{+0.40}$\\
Hercules&$133\pm12$ & $0.28$&$16.79_{-0.45}^{+0.45}$&$16.83_{-0.45}^{+0.45}$&$17.10_{-0.22}^{+0.22}$&$17.38_{-0.45}^{+0.45}$\\
Leo IV&$154\pm6$ & $0.16$&$16.56_{-0.90}^{+0.90}$&$16.64_{-0.90}^{+0.90}$&$16.68_{-0.45}^{+0.45}$&$17.22_{-0.90}^{+0.90}$\\
Leo V&$178\pm10$ & $0.07$&$16.82_{-0.70}^{+1.05}$&$16.94_{-0.72}^{+1.05}$&$16.35_{-0.35}^{+0.53}$&$17.23_{-0.70}^{+1.05}$\\
Leo T&$417\pm19$ & $0.08$&$17.28_{-0.37}^{+0.37}$&$17.32_{-0.37}^{+0.38}$&$16.70_{-0.19}^{+0.19}$&$17.35_{-0.37}^{+0.37}$\\
Segue 1&$23\pm2$ & $0.35$&$19.39_{-0.39}^{+0.39}$&$19.41_{-0.40}^{+0.39}$&$18.03_{-0.20}^{+0.20}$&$18.17_{-0.39}^{+0.39}$\\
Segue 2&$35\pm2$ & $0.19$&$17.06_{-1.75}^{+0.86}$&$17.11_{-1.76}^{+0.85}$&$16.64_{-0.87}^{+0.43}$&$17.08_{-1.75}^{+0.86}$\\
Ursa Major I&$97\pm4$ & $0.43$&$18.47_{-0.25}^{+0.25}$&$18.48_{-0.25}^{+0.25}$&$18.08_{-0.13}^{+0.13}$&$18.15_{-0.25}^{+0.25}$\\
Ursa Major II&$32\pm4$ & $0.53$&$19.38_{-0.39}^{+0.39}$&$19.38_{-0.39}^{+0.39}$&$18.51_{-0.19}^{+0.19}$&$18.48_{-0.39}^{+0.39}$\\
Willman 1&$38\pm7$ & $0.5$&$19.29_{-0.62}^{+0.91}$&$19.29_{-0.62}^{+0.91}$&$18.03_{-0.32}^{+0.45}$&$18.03_{-0.62}^{+0.91}$\\
Reticulum II&$30\pm3$ & $1.0$&$18.72_{-0.32}^{+0.85}$&$18.71_{-0.32}^{+0.84}$&$18.19_{-0.17}^{+0.42}$&$17.93_{-0.32}^{+0.85}$\\
Tucana II&$57\pm5$ & $1.0$&$19.10_{-0.58}^{+0.88}$&$19.05_{-0.58}^{+0.87}$&$18.79_{-0.29}^{+0.44}$&$18.45_{-0.58}^{+0.88}$\\
\hline
Horologium I&$79\pm7$ & $0.5$&$18.64_{-0.39}^{+0.95}$&-&$17.78_{-0.20}^{+0.47}$&-\\
Hydra II&$134\pm10$ & $0.5$&$16.56_{-1.85}^{+0.87}$&-&$16.89_{-0.92}^{+0.44}$&-\\
Pisces II&$182\pm18$ & $0.5$&$17.90_{-0.80}^{+1.14}$&-&$17.41_{-0.40}^{+0.57}$&-\\
Grus I&$120\pm11$ & $0.5$&$17.96_{-1.93}^{+0.90}$&-&$17.59_{-0.96}^{+0.46}$&-\\
\hline
\end{tabular}

\label{table:Jfs}
\end{table*}
\begin{figure}
$$\includegraphics[width=\columnwidth]{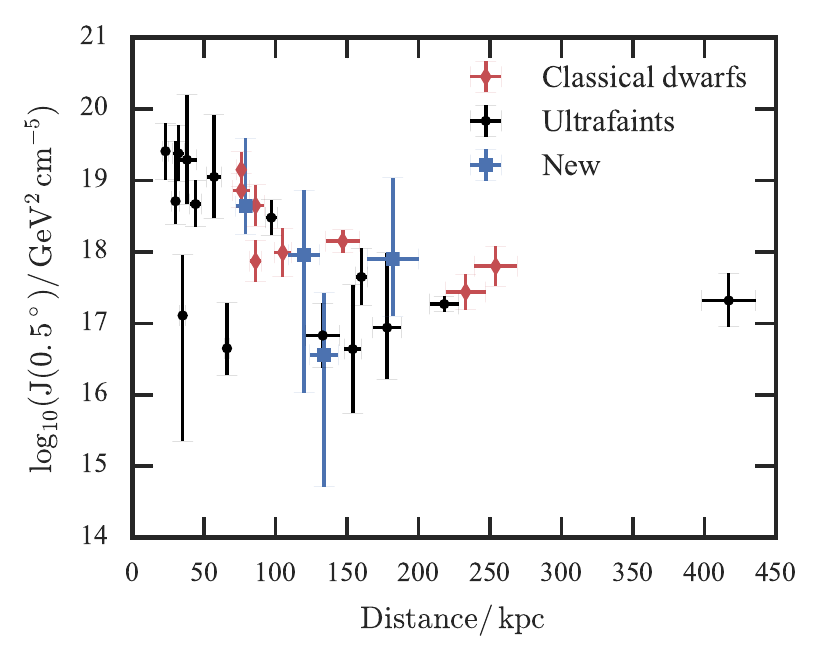}$$
\caption{Dwarf annihilation factors against distance: the points show
  the J-factors computed for an opening angle of $0.5^\circ$ using the
  simple NFW formula presented in this paper with $r_s=5R_\mathrm{h}$.
  The red diamonds correspond to the classical dwarfs, the black
  circles to the ultrafaints, and the blue squares to four
  recently-discovered dwarfs with no pre-existing literature estimates
  of their J-factors.}
\label{fig:J_against_d}
\end{figure}
\begin{figure*}
$$\includegraphics[width=\textwidth]{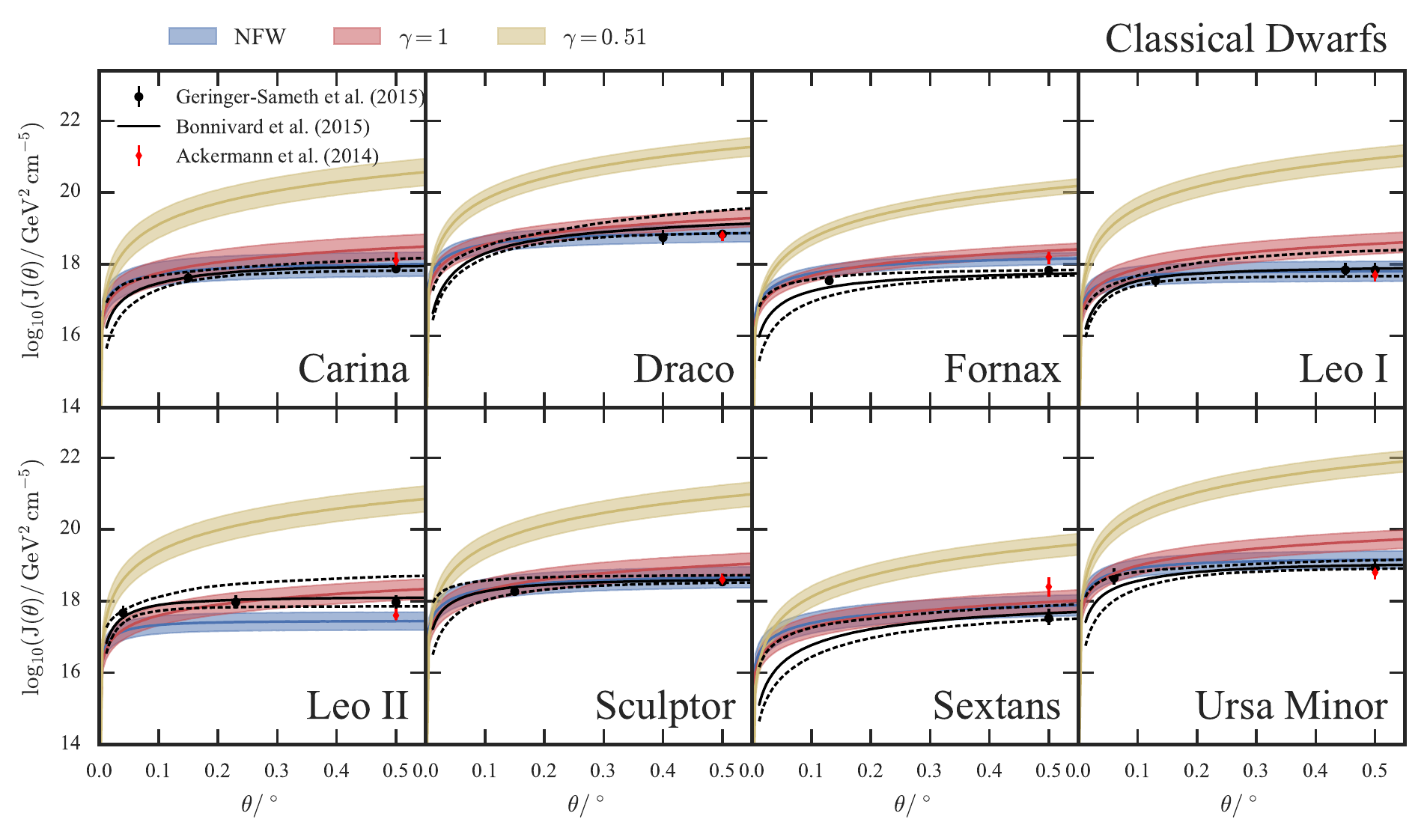}$$
\caption{Classical dwarf annihilation factors: the points are taken
  from the full Jeans analysis of \protect\cite{Ge15b}. The red and
  yellow bands are the estimates from equation~\eqref{eq:Jsimple}
  using $\gamma=1$ and $\gamma=0.51$ respectively. The blue band is
  the estimate from equation~\eqref{eq:Jfactorfull} using
  $r_s=5R_\mathrm{half}$. The median and $\pm1\sigma$ estimates of
  $\log_{10}(J(\theta))$ from \protect\cite{Bo15c} are given by the
  black solid and dashed lines respectively. The red diamonds are the
  estimates of $\log_{10}(J(0.5^\circ))$ from
  \protect\cite{FermiLAT}.}
\label{fig:cd_grid}
\end{figure*}
\begin{figure*}
$$\includegraphics[width=\textwidth]{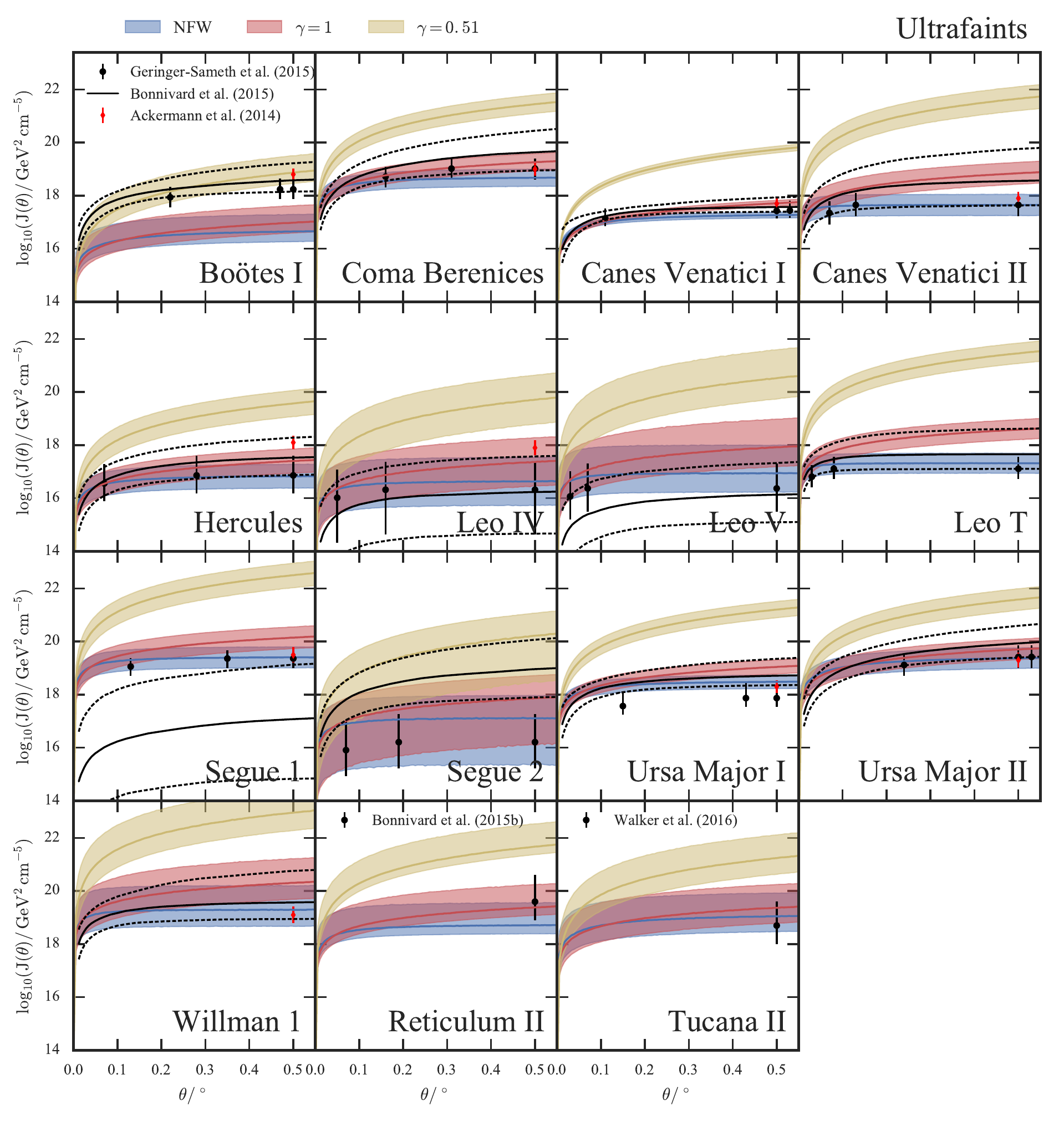}$$
\caption{Ultra-faint dwarf annihilation factors: see caption of
  Fig.~\protect\ref{fig:cd_grid}. The points for Tucana II and
  Reticulum II are taken from refs \cite{Wa15} and \cite{Bo15b}
  respectively.}
\label{fig:uf_grid}
\end{figure*}
\begin{figure*}
$$\includegraphics[width=\textwidth]{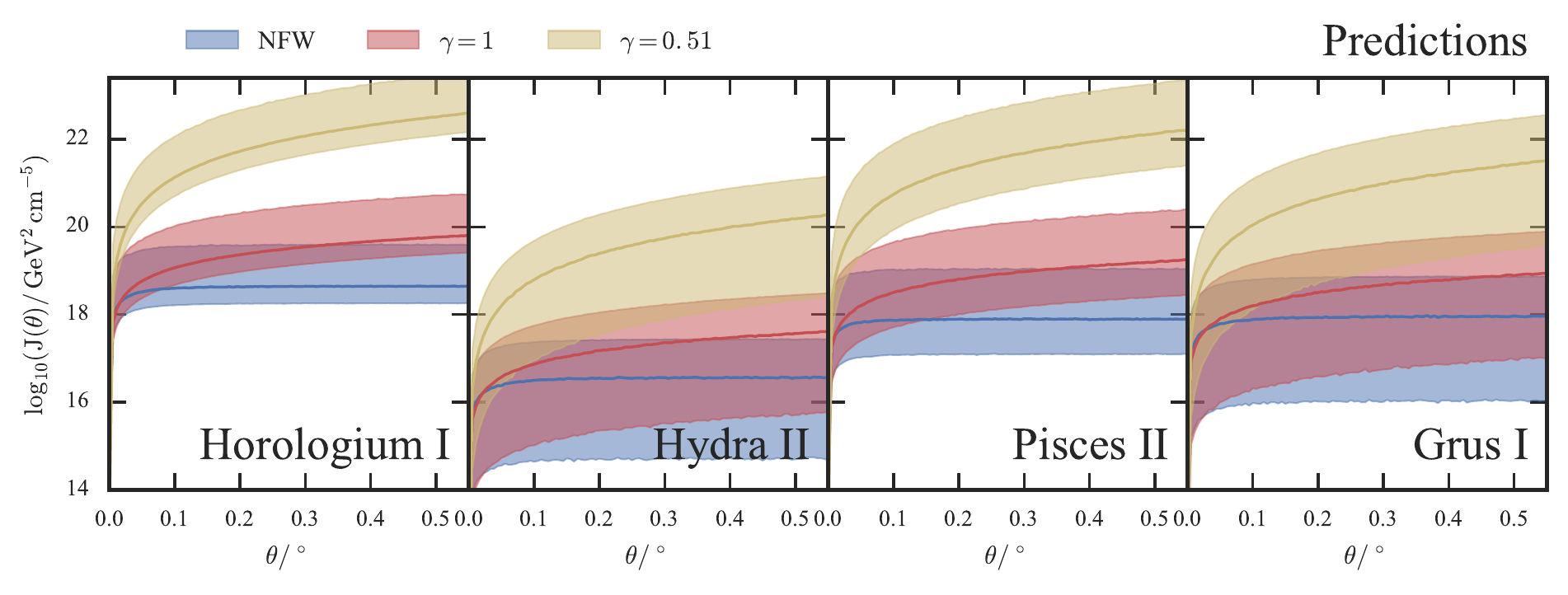}$$
\caption{Predictions for ultra-faint dwarf annihilation factors: see
  caption of Fig.~\protect\ref{fig:cd_grid}.}
\label{fig:predict_grid}
\end{figure*}

\section{Simple Formulae for J-factors}

Here, we provide a simple calculation for the J-factor of both a
spherical cusp and a Navarro-Frenk-White (NFW) halo. We show our
formulae reproduce the results of more sophisticated calculations.

\subsection{Spherical Cusp}

In spherical symmetry, dSphs follow the empirical law
\begin{equation}
\Mh = M(\rh) \approx {2.5 \over G} \sigmalos^2 \rh,
\label{eq:wolf}
\end{equation}
where $\rh$ is the (projected) half-light radius of the stars and
$\sigmalos$ is the luminosity weighted square of the velocity
dispersion. {\red Specifically, this is defined as
\begin{equation}
\sigmalos^2 = {1\over L} \int_V \nu(r) \vlos^2 dV
\end{equation}
where $L$ is the total luminosity, $\nu(r)$ is the luminosity density
of the dSph and $\vlos$ is the line of sight velocity. In other words,
the physical content of eq.~(\ref{eq:wolf}) is that the mass enclosed
within the half-light radius is robust against changes in
anisotropy. This result was demonstrated from solutions of the
spherical Jeans equations~\citep{Wa09,Wo10}. There are known biases in
this formula when applied to populations deeply embedded with dark
haloes, such as the metal-rich sub-populations of dSphs~\citep{WP}. In
our application to the entirety of the stellar population of the dSph,
any such bias is negligible, especially compared to the uncertainty in the
velocity dispersion itself, which dominates the overall error budget.}

Infinite spherical cusps obeying this law have enclosed mass
\begin{equation}
M(r) = \Mh \left( {r\over \rh} \right)^{3-\gamma}
=  {5 \sigmalos^2 \over 2 G \rh^{2-\gamma}} r^{3-\gamma},
\end{equation}
where $0<\gamma<3$.  The dark matter density is
\begin{equation}
\rhoDM(r) = {\Mh\over 4 \pi \rh^{3-\gamma}}{3 -\gamma \over r^\gamma}
= {5\sigmalos^2\over  8 \pi G \rh^{2-\gamma}}{3 -\gamma \over r^\gamma}.
\label{eq:sphericaldens}
\end{equation}
This gives us a one-parameter family of dark matter cusps that always
obey the empirical law~(\ref{eq:wolf}).

We now make two assumptions to enable us to perform the integration in
the J-factor analytically. First, we assume that the dSph is
sufficiently distant that
\begin{equation}
d\Omega d\ell \rightarrow {1\over D^2} 2 \pi R dR dz.
\end{equation}
Here, $D$ is the heliocentric distance, $z$ is the line of sight and
$R$ is a polar coordinate in the plane of the sky.  This approximation
means that the projection is from infinity rather than from a finite
distance. The integration volume is a cylinder rather than a cone. As
even the nearest ultrafaints (Segue 1 and Reticulum II) are $\sim
20-30$ kpc distant, this incurs little actual error.  Second, we
assume that the J-factor is dominated by contributions from the
singular cusp.  We show {\it a posteriori} that this approximation is
fine by comparing our formula to the results of more elaborate
calculations.

The integration in eq.~(\ref{eq:Jfactor}) can now be done
analytically
\begin{eqnarray}
\Jf &=& {2\pi\over D^2} \int_{-\infty}^{\infty} dz \int_0^{D\angle} R \rhoDM^2 dR, \nonumber \\
%&=& {\sigmalos^4 \over G^2}{ D^{1-2\gamma} \angle^{3-2\gamma}\over \rh^{4 -2 \gamma}} \times P(\gamma), \nonumber \\
&=& {25 \sigmalos^4 \over 64 G^2} {1 \over D^2 \rh} \left({D\angle \over \rh}\right)^{3-2\gamma} P(\gamma),
\label{eq:Jsimple}
\end{eqnarray}
where $P(\gamma)$ is a dimensionless number depending on the cusp
slope $\gamma$
\begin{equation}
P(\gamma) = {2 \over \pi^{1/2}}\,{(3-\gamma)^2\Gamma(\gamma -
  \frac{1}{2})\over (3-2\gamma)\Gamma(\gamma)}.
\end{equation}
{\red Here, $\Gamma(x)$ is the gamma function, whilst we require $1/2
  < \gamma < 3/2$ for convergence. This constraint can be interpreted
  physically. If the density is too shallow ($\gamma < 1/2$), then the
  dark matter extends too far along the line of sight, causing the $z$
  integral to diverge.  Similarly, if it is too strongly cusped
  ($\gamma> 3/2$), then the contribution from the cusp at $r=0$ also
  causes the $z$ integral to diverge.}

The angular integration is usually performed out to $\angle =
0.5^\circ$ from the centre of the dSph, as this is typical of the
resolution of Fermi-LAT data in the GeV range.  \citet{Wa11} have
argued that the J-factor is most robust to modelling uncertainty when
the integration angle is $\acrit \approx 2\rh/D$, or twice the
half-light radius of the stars divided by distance to the dSph.
Finally, it is also useful to compute the J-factor out to $\amax ={\rm
  asin} (r_\mathrm{max}/D)$ where $r_\mathrm{max}$ is an estimate of
the maximum galactocentric distance in the sample of observed member
stars (see~\citep[][Sec. 6.2]{Ge15b}).  We expect our formula to break
down at large $\angle$, but this is usually comparable or beyond
$\amax$.

In scenarios in which the dark matter decays (rather than
annihilates) to give $\gamma$ rays~\citep{Bo10}, it is also
helpful to have the D-factor, which is just
\begin{equation}
\Df = \int\int \rhoDM (\ell, \Omega) d\ell d\Omega.
\label{eq:Dfactor}
\end{equation}
Using the same approximation of an infinite cusp obeying the empirical
law (\ref{eq:wolf}), we find
\begin{equation}
\Df = {5 \sigmalos^2 \over 8 G} {\rh \over D^2} \left( {D\angle \over \rh}\right)^{3-\gamma} Q(\gamma),
\label{eq:Dsimple}
\end{equation}
where
\begin{equation}
Q(\gamma) = {2 \pi^{1/2}}\,{\Gamma( \frac{\gamma}{2} - \frac{1}{2}) \over \Gamma(\gamma/2)}.
\end{equation}
with $1 < \gamma < 3$ for convergence. This of course does not
converge for $\gamma =1$ because of contributions at large radii where
the integral diverges logarithmically.

\begin{figure*}
$$\includegraphics[width=\textwidth]{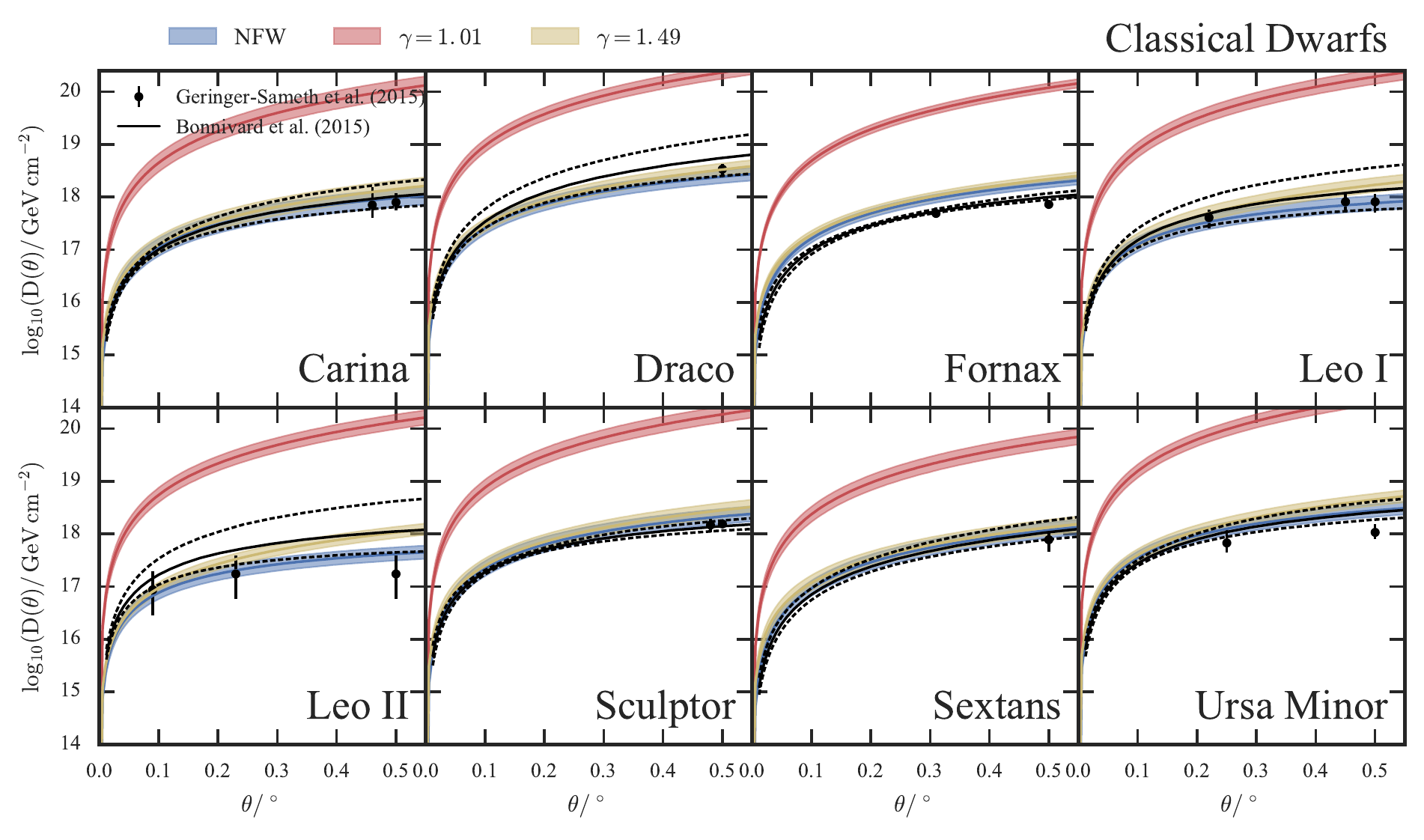}$$
\caption{Classical dwarfs decay factors: the points are taken from the
  full Jeans analysis of \protect\cite{Ge15b}. The red and yellow
  bands are the estimates from equation~\eqref{eq:Dsimple} using
  $\gamma=1$ and $\gamma=1.49$ respectively. The blue band is the
  estimate from equation~\eqref{eq:Dfactorfull} using
  $r_s=5R_\mathrm{half}$. The median and $\pm1\sigma$ estimates of
  $\log_{10}(D(\theta))$ from \protect\cite{Bo15c} are given by the
  black solid and dashed lines respectively.}
\label{fig:cd_grid_decay}
\end{figure*}

\subsection{NFW Cusp}

There is ample numerical evidence that dark matter halos have an
approximate double power law structure~\citep[][hereafter NFW]{NFW}
with cusp slope $\gamma = 1$.  For this astrophysically important
case, we obtain
\begin{equation}
\Jf = {25 \over 8 G^2}\, {\sigmalos^4 \angle \over D \rh^2},
\label{eq:JsimpleNFW}
\end{equation}
This very simple result does not appear to have been given before.
The D-factor however is infinite, as the surface density of an
untruncated $1/r$ density cusp does not converge. In fact, for the
full NFW model, it is possible to carry out the integration explicitly
for both the J- and the D-factors and obtain exact results.  These are
a little more cumbersome than the pure power-law case, but still
simple enough to be useful.

We begin by introducing an auxiliary function \citep[e.g.,][]{He90,Ev14}
\begin{equation}
X(s) = \begin{cases}
{\displaystyle {1\over \sqrt{1-s^2}}\arcsech\, s}, & 0\le s\le 1,\nonumber\\
{\displaystyle {1\over \sqrt{s^2-1}}\arcsec\, s }, & s\ge 1.
\end{cases}
\end{equation}
We note that $X(1)=1$ so that the function is continuous.  We take the
NFW model in the form
\begin{equation}
\rhoDM(r) = {\rho_0 \rs^3 \over r ( r + \rs)^2}.
\end{equation}
Then, the J-factor is
\begin{eqnarray}
\Jf &=& {\pi \rho_0^2 \rs^2 \over 3D^2\Delta^4}\Biggl[ 2D\angle (7D\rs^3\angle - 4D^3\rs\angle^3 + 3\pi \Delta^4) \cr
&+& {6\over \rs} (2\Delta^6 - 2\rs^4\Delta^2 - D^4\rs^2\angle^4) X\Bigl(
  {D\angle\over \rs}\Bigr)\Biggr],
\label{eq:Jfactorfull}
\end{eqnarray}

Then, defining $y = D\angle/\rs$, the J-factor is
\begin{eqnarray}
\Jf &=& {\pi \rho_0^2 \rs^3 \over 3D^2\Delta^4}
\Biggl[ 2 y (7y - 4y^3 + 3\pi \Delta^4) \cr
&+& 6 (2\Delta^6 - 2\Delta^2 - y^4) X(y)\Biggr],
\label{eq:Jfactorfull}
\end{eqnarray}
where $\Delta^2 = 1- y^2 = 1- D^2\angle^2/\rs^2$.  Given a mass $\Mh$ enclosed
within the half-light radius $\rh$ the parameter $\rho_0$ is given by
\begin{equation}
\rho_0 = \frac{\Mh}{4\pi\rs^3}\Big(\log\Big[\frac{\rs+\rh}{\rs}\Big]-\frac{\rh}{\rs+\rh}\Big)^{-1}.
\label{eq:nfwhalorho0}
\end{equation}
On identifying $\rho_0 \rs = 5\sigmalos^2/(4\pi \rh G)$ and letting
$\rs \rightarrow \infty$, we obtain the infinite cusp
(\ref{eq:sphericaldens}) with $\gamma=1$.  In this limit, the full
J-factor (\ref{eq:Jfactorfull}) reduces to (\ref{eq:JsimpleNFW}), as
it should. In the other limit, as $\angle$ becomes large, the J-factor
curve turns over and tends to the asymptotic value
\begin{equation}
\Jf \rightarrow {4 \pi\over 3} {\rho_0^2 \rs^3 \over D^2}
\label{eq:nfwhaloasymp}
\end{equation}
The D-factor is
\begin{equation}
\Df = {4\pi \rho_0 \rs^3 \over D^2}\Biggl[ \log \left(
  {D\angle \over 2\rs} \right) +X\Bigl(
  {D\angle\over \rs}\Bigr)\Biggr]
\label{eq:Dfactorfull}
\end{equation}
with the limit of an infinite cusp $\rs \rightarrow \infty$
regenerating the result of (\ref{eq:Dsimple}) with $\gamma =1$.  In
particular, as $\angle$ becomes large, the first term in
eq~(\ref{eq:Dfactorfull}) dominates and the D-factor diverges
logarithmically. However, as $\angle \rightarrow 0$, the D-factor
tends to
\begin{equation}
\Df \rightarrow 2\pi \rho_0 \rs \angle^2 \log \left( {D\angle \over 2\rs} \right)
\end{equation}
and so approaches zero quadratically.

\begin{figure*}
$$\includegraphics[width=\textwidth]{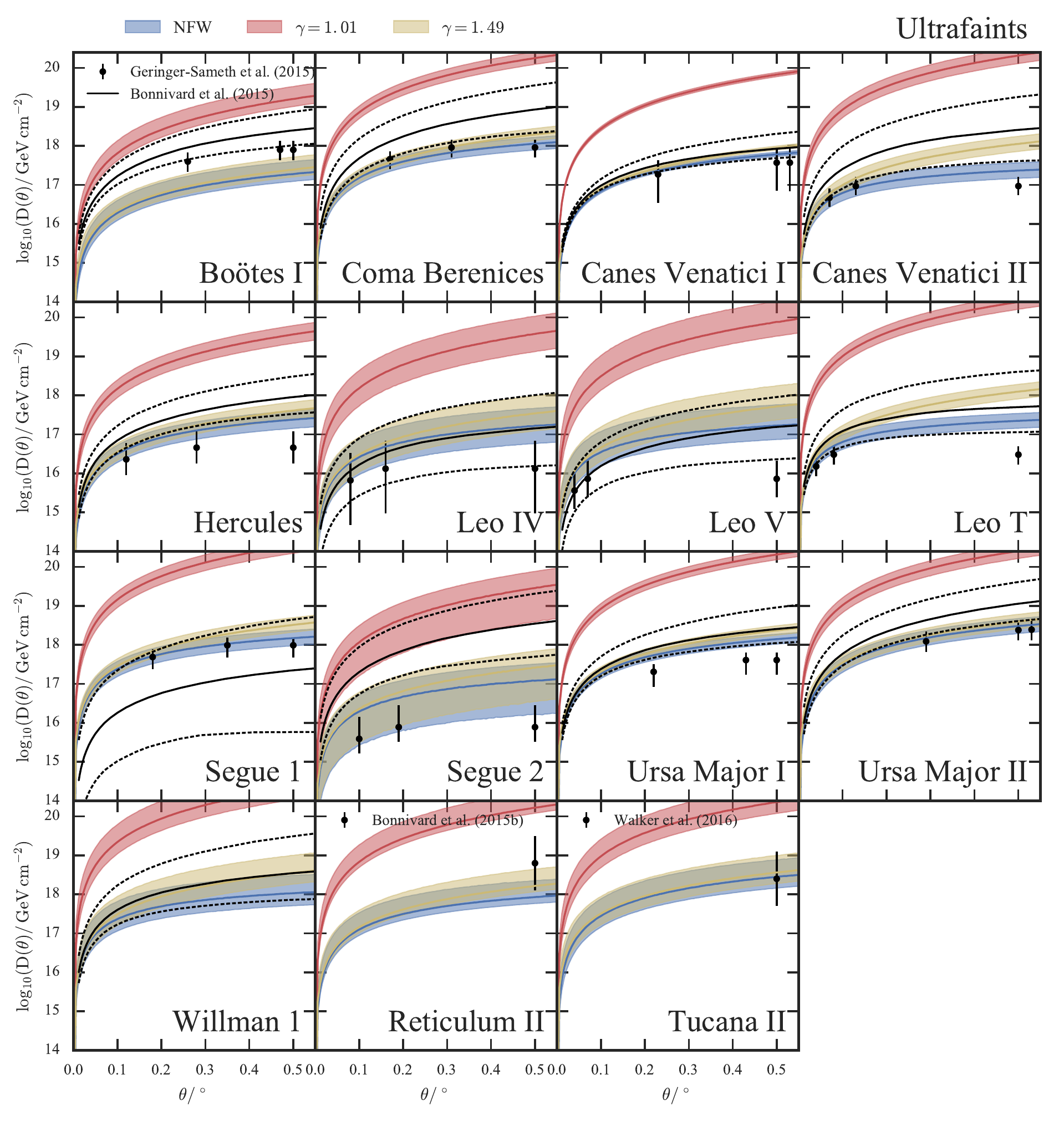}$$
\caption{Ultra-faint dwarfs decay factors: see caption of
  Fig.~\protect\ref{fig:cd_grid_decay}.}
\label{fig:uf_grid_decay}
\end{figure*}
\begin{figure*}
$$\includegraphics[width=\textwidth]{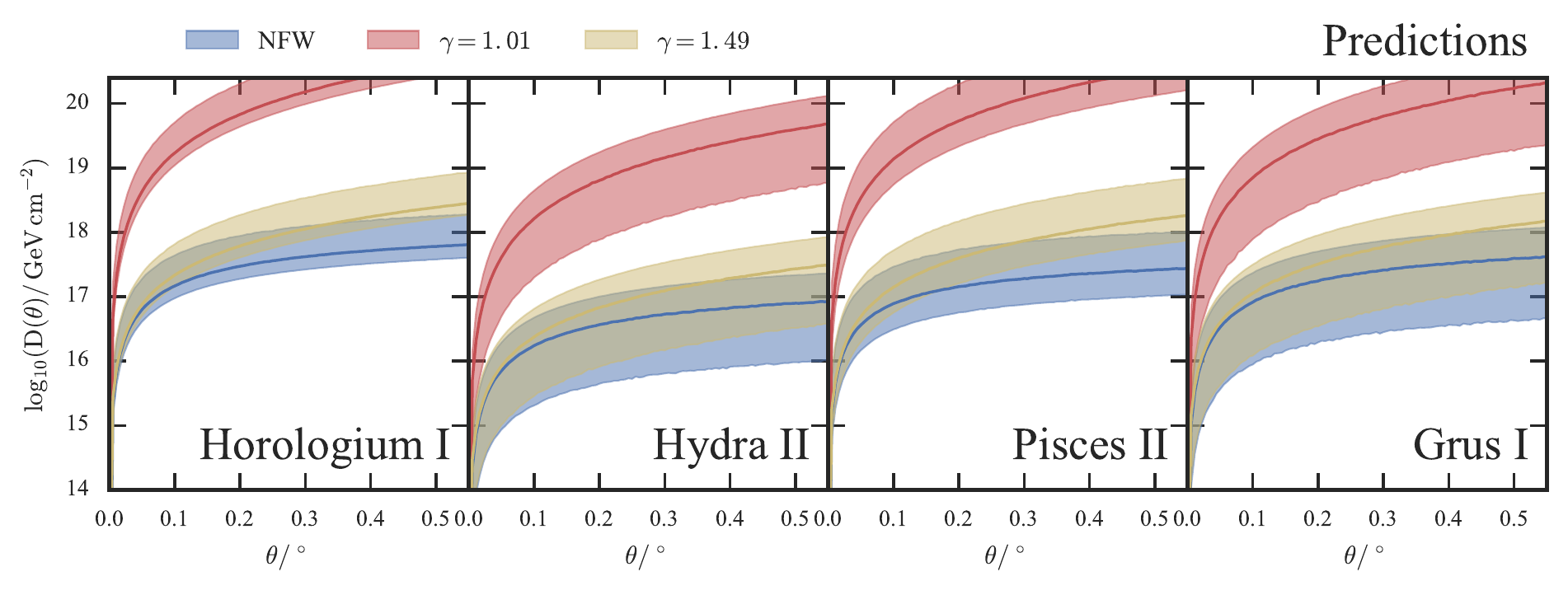}$$
\caption{Predictions for ultra-faint dwarf decay factors: see caption
  of Fig.~\protect\ref{fig:cd_grid_decay}.}
\label{fig:predict_grid_decay}
\end{figure*}

\subsection{Comparisons}

Here, we compare our formulae with results in the
literature. J-factors are often computed by finding the dark matter
density profile that best fits the stellar kinematics through the
spherical Jeans equations~\citep{Wa11,Bo15a,Ge15b}. The
multi-dimensional likelihood functions, which usually involve
parameters controlling the dark halo structure as well as the velocity
anisotropy, are then explored with Monte Carlo methods based on
Bayesian parameter inference. An alternative method~\citep{Mar15},
which was used by the Fermi-LAT colalboration, has some points of
contact with our approach here. They use the empirical relationship
~(\ref{eq:wolf}) between velocity dispersion and mass enclosed within
the half-light radius to circumvent solution of the Jeans
equation. They then construct likelihood functions for each dSph from
the luminosity, half-light radius and mass enclosed, together with
priors, which are explored with a two-level Bayesian hierarchical
model. Once the halo parameters are derived, the NFW profile is
numerically integrated to find the J-factor.

The J-factors and D-factors computed using the full formula for an NFW
cusp for a range of dSphs are given in Table~\ref{table:Jfs}.  Error
bars are computed via propagation of errors in velocity dispersion,
distance, and halflight radius using Eq.~\ref{eq:Jfactorfull}.  {\red
  We use an `ellipticity-corrected' version of the half-light radius
  which is given by the geometric average of the half-light radius as
  measured along the major and minor axes. This amounts to multiplying
  by $\sqrt{1-\epsilon}$ where $\epsilon$ is the ellipticity. Most
  data on half-light radii, distances, ellipticities and central
  velocity dispersions are extracted from the recent compilation of
  ~\citep{Mc12}. In addition, we use data from
  \cite{Ko15,Be10,Ma15,Ki13,Ki15,Wa15} on more recently-discovered
  dwarfs. In cases where the ellipticity is only bounded we use the
  reported upper-bound for our `ellipticity-corrected' half-light
  radius. }  There is very little difference between values in this
table and ones inferred from a Jeans analysis. We also show the
J-factors for an opening angle of $\theta=0.5^\circ$ against distance
for the dSphs in Fig.~\ref{fig:J_against_d}. Naturally, the J-factors
decrease with distance making the closest dSphs the most attractive
candidates for dark matter annihilation detection. However, the
diversity in measured velocity dispersions and half-light radii
introduce variation into this J-factor versus distance relation.

In Figs.~\ref{fig:cd_grid} and~\ref{fig:uf_grid}, we show how our
simple formulae compare to the results of Jeans calculations for the
J-factor profiles for the classical and the ultrafaint dwarfs
respectively. Red and yellow lines show the results obtained using
eq.~(\ref{eq:Jsimple}). In each case, coloured bands show the
1$\sigma$ range of values obtained by error propagation using the
reported error on the data (half-light radius, line of sight velocity
dispersion and heliocentric distance)\footnote{In the cases of Segue
  2, Hydra I and Grus I only 95\percent\, upper bounds, $C$, on the
  line-of-sight velocity dispersion are available. For these we assume
  a uniform distribution on the line-of-sight velocity dispersion
  between $0.1\kms$ and $C/0.95$.} The blue bands use
eq.~(\ref{eq:Jfactorfull}) with $\rs = 5\rh$, which is a reasonable
summary of results from phase-space modeling of the dSphs, as given in
Figure 8 of ref~\citep{Am11}.  For comparison, computed values for the
J- and D-factors from \citep{Ge15b} are also shown as solid circles
with error bars.  They are derived by assuming parametric laws for the
light profile, the anisotropy of the stars and the dark halo
profile. The latter is permitted to have a general double power-law
structure with the density falling like $r^{-\gamma}$ at small radii
and $r^{-\beta}$ at large radii. Solution of the spherical Jeans
equations and subsequent projection then provides the line of sight
velocity dispersions. By choosing priors on the unknown parameters and
sampling the likelihood function through Markov Chain Monte Carlo
Methods, \citet{Ge15b} obtain constraints on the dark matter
distribution. Hence, they can calculate the median value of the
J-factors and the $1\sigma$ distribution at selected angles (see their
Table 2). The data points show the results of their calculations at
two locations, namely $\angle_{1/2}$ or the angle containing 50 per
cent of the emission, and $\angle_{\rm max} = {\rm arcsin} (r_{\rm
  max})/D$ where $r_{\rm max}$ is an estimate of the distance to the
outermost member star with a measured radial velocity. We also show
the J-factor profiles with $\pm1\sigma$ errors computed by
\citet{Bo15c}, which were obtained through a Jeans analysis similar to
\citep{Ge15b}. Finally, we show the constraints on the J-factors at
$\theta=0.5^\circ$ from the Fermi-LAT collaboration \citep{FermiLAT}.

We see that the simple formula~(\ref{eq:Jsimple}) performs reasonably
well for the classical dwarfs, and better still for the
ultrafaints. At larger $\angle$, the J-factor for pure power-law cusps
is overestimated -- here the assumption of an infinitely extended cusp
breaks down.  The full formula for the NFW halo (\ref{eq:Jfactorfull})
removes even this deficiency and reproduces the computational results
extremely well.  Note that it is now just as straightforward to work
out the entire profile as to compute a single value. In principle, the
full profile may give information on the structure of the dark halo if
a dSph is detected in $\gamma$ rays.  The largest discrepancy between
our calculation and that of \citet{Ge15b} is for Bo\"otes I. {\red
  This may arise because Bo\"otes I possesses both a hot and cold
  component~\citep{Ko11}, whereas our calculation uses the velocity
  dispersion of the colder population only}.  Another possible source
of this discrepancy is that the best-fitting dark halo density law
inferred by \citep{Ge15b} may deviate from the strict NFW form -- in
fact, they report that Bo\"otes I has a median value of its inner
slope $\gamma$ of 0.53 and outer slope $\beta$ of 5.9.  This
discrepancy persists when comparing against the results of
\citet{Bo15c}. The other striking anomaly is when comparing our result
for Segue 1 with \citet{Bo15c}.  This is caused by different ways of
designating a subset of the spectroscopically observed stars as
``members'' of Segue 1 as opposed to Milky Way foreground
contaminants, as originally noted in \citep{NO09}. \citet{Bo15b} have
shown that the determination of Segue 1's J-factor (and, implicitly,
its velocity dispersion) is extremely sensitive to the inclusion or
exclusion of a small number of stars with ambiguous membership
status. The J-factors computed in~\citep{Bo15c} are based on more
stringent membership criteria for Segue 1 as compared to the analyses
of~\citep{Ge15b} and~\citep{FermiLAT}.

A straightforward conclusion from Figs.~\ref{fig:cd_grid}
and~\ref{fig:uf_grid} is that Draco, Ursa Minor, Sculptor, Coma
Berenices, Reticulum II, Tucana II and Ursa Major II are the most
favourable dSphs for which to look for signatures of dark matter
annihiliation. Willman 1 also has a high J-factor but the assumption
of dynamical equilibrium for this object is dubious as it appears to
be severely tidally disrupted \citep{Willman1}. Similarly, Segue 1 has
a high J-factor but the issue of foreground contamination of the
spectroscopic sample brings this into doubt \citep{NO09,Bo15b}.

Fig.~\ref{fig:predict_grid} gives predictions for 4 recently
discovered ultrafaints for which there are no J-factors in the
literature. These objects are Horologium I and Grus I (discovered in
{\it Dark Energy Survey Data} \citep{Ko15,Be15}, Hydra II (discovered
by the {\it Survey of the Magellanic Stellar History}) and
\citep{Ma15}, Pisces II (discovered in {\it Sloan Digital Sky Survey}
data \citep{Be10}).  Keck/DEIMOS spectroscopy of Hydra II and Pisces
II has recently been published \citep{Ki15}.  The Gaia-ESO survey has
measured the velocities of 5 stars in Horologium I using the
VLT/Giraffe combination~\citep{Ko15a}, whilst Magellan/M2FS has been
used to target 7 stars in Grus I~\citep{Wa15}. Using the data in these
papers, we calculate the J-factors.  The uncertainties are calculated
by Monte Carlo sampling and then estimating the $1\sigma$ bound on the
J-factor from the ensuing distributions.  It is clear that Horologium
I is another excellent candidate with a J-factor comparable to
Reticulum~II and Tucana II.

We show in Figs~(\ref{fig:cd_grid_decay} -
\ref{fig:predict_grid_decay}) analogous plots for the D-factor for
classical dwarfs, ultrafaints and predictions for recent discoveries.
Both the simple formula (\ref{eq:Dsimple}) and the exact result for an
NFW cusp (\ref{eq:Dfactorfull}) do an excellent job of reproducing the
results in the literature -- with much greater economy of effort!

\section{Applications}

\subsection{Cusps and Cores}

So far, we have focussed on the NFW model with its $1/r$ density cusp.
Although this has a preferred status because of its importance in
theories of galaxy formation, there is some evidence that dark haloes
may have a different structure. For example, many of the detailed
kinematical fits reported by \citep{Ge15b} have milder cusps, often
with $\rhoDM \sim r^{-1/2}$ at small radii.  There is also strong
observational evidence that some of the dwarf galaxies are mildly
cusped or even cored. This includes the persistence of substructure in
Ursa Minor~\citep{Kl03}, the survival of globular clusters in
Fornax~\citep{Co12}, and the kinematics of multiple
populations~\citep{WP,Am12,Ag12,Am13}. Here, somewhat speculatively,
we suppose that the J-profile can be mapped out as a function of
$\angle$ and ask what can then be deduced about the dark halo
structure.

A flexible set of dark halos with double-power law structure has the
form~\citep{Zh96,Ch11}
\begin{equation}
\rhoDM(r) = \rho_0
\Big(\frac{r}{\rs}\Big)^{-\gamma}\Big(1+\Big(\frac{r}{\rs}\Big)^{\alpha}\Big)^{(\gamma-\beta)/\alpha},
\label{eq:zhao}
\end{equation}
with $\alpha,\beta$ and $\gamma$ as positive numbers.  The familiar
NFW profile is recovered in the case $(\angle,\beta,\gamma) =
(1,3,1)$.  Although the J-factors for this entire class of models are
not analytic, nonetheless we can easily work out their general
properties. At small radii, the density is cusped like $r^{-\gamma}$.
For strong cusps with $1/2 < \gamma < 3/2$, the behaviour can be deduced
from our work on infinite cusps in Section II by making the
identification $\rho_0 r_s^\gamma = (3-\gamma)\sigmalos^2/(\pi
\rh^{2-\gamma} G)$. We see from Eq.~\ref{eq:Jsimple} that the J-factor
increases like $J \propto \angle^{3-2\gamma}$ on moving outwards from
the center of the halo. For weaker cusps and cores ($0 \le \gamma
<1/2$), the dominant term in the J-factor near the origin is always a
gentler quadratic, namely $J \propto \angle^2$, as the $\gamma$-ray
emission is now no longer dominated by the very centre.  Beyond the
scale radius $\rs$, the J-factor turns over and tends to a constant
value. This is given by
\begin{equation}
\Jf \rightarrow {4\pi \rs^3\rho_0^2 \over D^2}
{\Gamma[(2\beta-3)/\alpha]\Gamma[(3-2\gamma)/\alpha]\over \alpha
  \Gamma[2(\beta-\gamma)/\alpha]}.
\label{eq:Jasymptotic}
\end{equation}
Formally, we require $\beta>3/2$ and $\gamma<3/2$ for convergence,
though this is satisfied by almost all astrophysically realistic
models.  Notice when $(\alpha,\beta,\gamma) = (1,3,1)$, we recover the
asymptotic value for the J-factor of an NFW halo
(eq.~\ref{eq:nfwhaloasymp}).

The behaviour of the D-factor can be worked out in a similar way,
though now the boundary between strong and weak cusps is at
$\gamma=1$. For cusps with $\gamma <1$, the D-factor initially rises
like $J \propto \angle^{3-\gamma}$, whilst for cusps with $\gamma \le
1$, the rise is quadratic $J \propto \angle^2$. At large angles, the
D-factor turns over to the constant value
\begin{equation}
\Df \rightarrow {4\pi \rs^3\rho_0 \over D^2}
{\Gamma[(\beta-3)/\alpha]\Gamma[(3-\gamma)/\alpha]\over \alpha
  \Gamma[(\beta-\gamma)/\alpha]},
\label{eq:Dasymptotic}
\end{equation}
provide $\beta>3$ and $\gamma<3$, as required for convergence. For the
NFW model, $\beta =3$ and so the D-factor logarithmically diverges.

We can illustrate this behaviour with two simple models.  A prototype
of a weakly cusped dark halo is the model with $(\alpha,\beta,\gamma)
= (1,3,1/2)$, namely:
\begin{equation}
\rho(r) = {\rho_0 \rs^3 \over r^{1/2}(r + \rs)^{5/2}}
\end{equation}
for which the J-factor is
\begin{eqnarray}
\Jf &=& {\pi \rho_0^2 \over 6 \rs^3 \Delta^6}\Biggl[ 6D^2\rs^2\angle^2
  -2D^4\angle^4 -19\rs^4\cr
&+& 3\rs^2(4\rs^2 + D^2\angle^2)X\Bigl(
  {D\angle\over \rs}\Bigr)\Biggr].
\label{eq:Jfactorweak}
\end{eqnarray}
Its asymptotic value is
\begin{equation}
\Jf \rightarrow {\pi \rs^3\rho_0^2 \over 3 D^2}
\end{equation}
while its central value is
\begin{equation}
\Jf \rightarrow \frac{\pi}{6} \rs \rho_0^2\left( 6\log\left(
  {4\rs^2\over D^2\angle^2}\right) -19\right) \angle^2
\end{equation}
The D-factor is not analytic, but diverges logarithmically at large
radii.

A prototype for cored dark haloes is the famous Plummer (1911) model,
which corresponds to $(\alpha,\beta,\gamma) = (2,5,0)$. This is often
used for modelling cored dark haloes (as well as clusters and the
stellar populations in dSphs). The density is
\begin{equation}
\rho(r) = {\rho_0 \rs^5 \over (r^2 + \rs^2)^{5/2}}.
\end{equation}
The J-factor is
\begin{equation}
\Jf = {5\pi^2\over 64}{\rho_0^2 \rs^3\over D^2} \left[ 1- {\rs^7\over
      (\rs^2 + D^2\angle^2)^{7\over2}}\right].
\end{equation}
Notice that the J-factor behaves like $\theta^2$ at small angles,
whereas the asymptotic value is in accord with our general rule
(\ref{eq:Jasymptotic}).  The D-factor is also very simple
\begin{equation}
\Df = {4\pi\over 3} {\rho_0\rs^3\angle^2 \over (\rs^2 + D^2\angle^2)}.
\end{equation}
It again approaches zero quadratically, and tends to the asymptotic
value given by (\ref{eq:Dasymptotic}). This is in accord with our
general results.

In principle, if the variation of the J-factor can be mapped out with
integration angle, then valuable information on the structure of the dark
halo can be gleaned. The behaviour at small angles can yield
information on the cusp slope $\gamma$. In particular, if $\gamma >
1/2$, then the logarithmic gradient of $J$ with respect to $\angle$ is
the cusp slope. Alternatively, if $J \propto \angle^2$, then the cusp
is either very mild or the dark halo is cored. Similarly, the
asymptotic value of the J-factor is controlled by the normalisation
$\rho_0$, scale-length $\rs$ and the density fall-off $\beta$. Some of
these quantities, such as $\rs$ can of course be constrained by the
stellar kinematics of the stars, so the asymptotic value may enable a
complete solution for the halo to be obtained.

\begin{figure}
$$\includegraphics[width=0.5\textwidth]{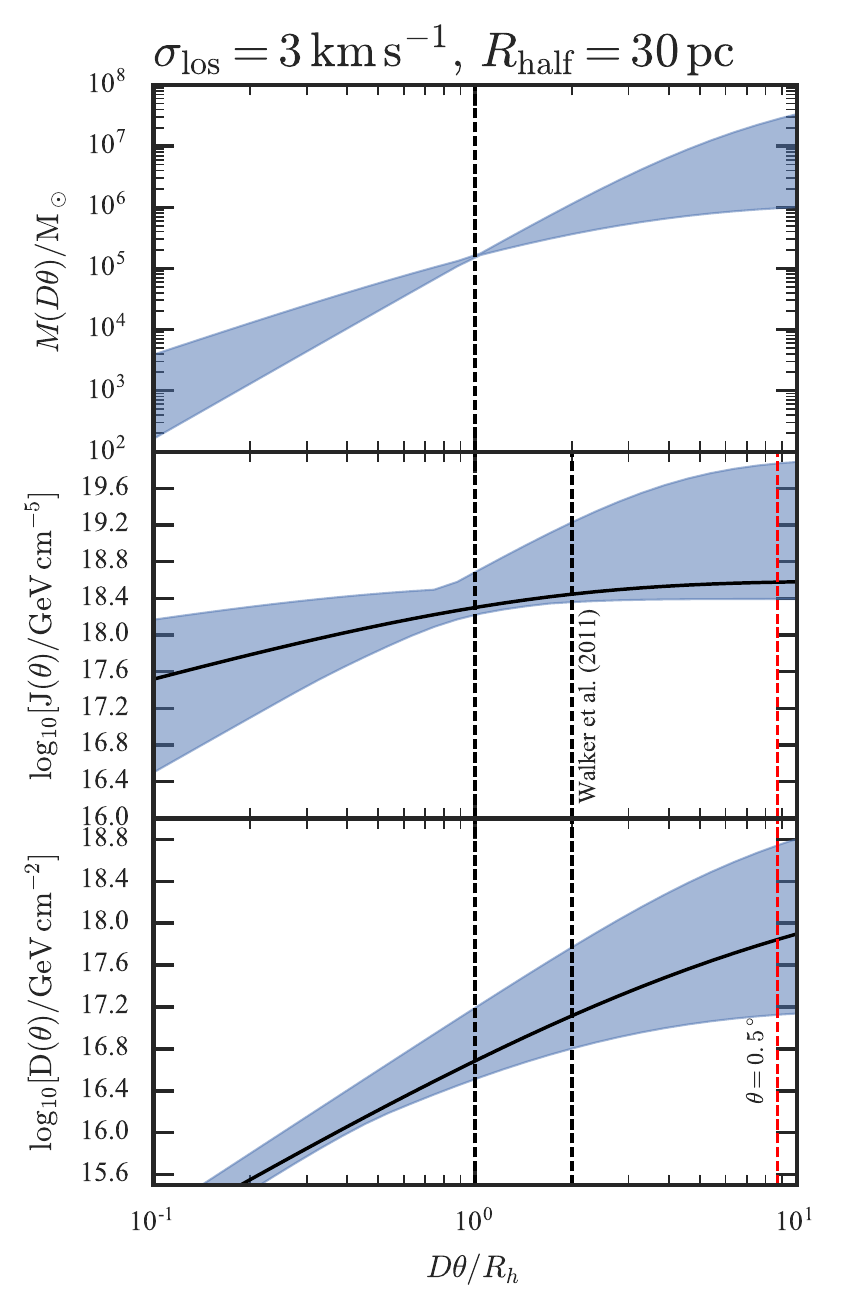}$$
\caption{Comparison of J-factors and D-factors of
  ($\alpha,\beta,\gamma$) dark halos with the same mass enclosed
  within the half-light radius ($\rh=30\,\mathrm{pc}$), represented by
  the left black dashed line. The parameters chosen approximately
  match those of Reticulum II and we set $r_s=5\rh$. The blue band
  shows the range of possible profiles when
  $\alpha\in[1.,2.],\beta\in[3.,6.],\gamma\in[0.,1.2]$. The solid
  black line corresponds to the NFW profile. Notice that the J-factors
  of these models show least scatter at the angle $\angle \approx
  \rh/D$. The radii suggested in \citep{Wa11} is indicated by a
  vertical dotted line, and -- at least for these models -- shows
  greater scatter. The D-factors show the smallest scatter at a much
  smaller angle. The red dashed vertical line shows Fermi-LAT
  resolution for Reticulum II.}
\label{fig:sweetspot}
\end{figure}

\subsection{The Sweet Spot}

The idea that there is a sweet spot -- a location in which the
measured quantity is reasonably robust against changes in the
underlying dark halo model -- has proved quite powerful in studies of
dSphs. Of course, the most successful instance is the hypothesis that
the mass of dark matter enclosed within the half-light radius of the
stellar population is reasonably robust against changes in
anisotropy~\citep{Wa09,Wo10}. Similar ideas have been used to identify
radii at which the enclosed mass is insensitive to changes in the
underlying halo models~\citep{Am11, Ha16}. It is therefore natural to
wonder if there is a special angle $\angle$ at which the J-factor is
particularly robust.

\citet{Wa11} noticed from their Jeans solutions for dark matter haloes
for the Carina dSph that the J-factor was least sensitive to changes
in the inner slope $\gamma$, outer slope $\beta$ and velocity
anisotropy at an integration angle $\angle \approx 2\rh/D$. They then
generated mock datasets from distribution functions for models with
varying inner slope $\gamma$ and velocity anisotropy, but fixed outer
slope $\beta$, which supported the idea of a sweet spot at $\angle
\approx 2\rh/D$.

We test this directly in Fig.~\ref{fig:sweetspot} by plotting the
range of J-factors and D-factors of ($\alpha,\beta,\gamma$) models
with the same mass enclosed within the half-light radius of the stars
$\rh$. This mass was chosen using the velocity dispersion and
half-light radius of Reticulum II. The range of models explored were
$\alpha\in[1.,2.],\beta\in[3.,6.],\gamma\in[0.,1.2]$. Although we have
not explicitly used any kinematic data, it is nonethless encoded in
the models using the fact that the mass enclosed with $\rh$ -- and
hence the luminosity weighted velocity dispersion -- is the same for
all the models~\citep{Wa09,Wo10}. The vertical line in
Fig.~\ref{fig:sweetspot} shows that the integration angle $\angle
\approx \rh/D$ at which the scatter in the J-factors of the models is
minimized. Notice even though the mass enclosed within $\rh$ is
exactly the same for all the models, there is nonetheless some scatter
in the J-factors even at $\angle = \rh/D$. Also shown with a vertical
line is the angle $\angle \approx 2\rh/D$ suggested by \citep{Wa11},
at which the scatter is rather greater.

Presumably, the explanation of this discrepancy is that the models
constructed in \citep{Wa11}, which are derived from Markov Chain Monte
Carlo fits to discrete radial velocities, do not exactly satisfy the
empirical relation on mass enclosed within the half-light radius and
velocity dispersion. Of course, the actual integration angle used is a
trade-off between gathering power and minimizing the uncertainty in the
dark halo properties.

\section{Conclusions}

The main advance in this paper is the provision of very simple
formulae for J-factors for indirect dark matter detection. This
includes power-law spherical cusps, as well as the exact result for
the Navarro-Frenk-White~\citep{NFW} model. These compact formula offer
significant savings in effort over Jeans solution methods driven by
Markov Chain Monte Carlo engines.  Especially for the ultrafaint
dwarfs, for which the size of the datasets of discrete velocities is
tiny, this provides an entirely elementary method of evaluating
J-factors. {\red The exact formulae for the J-factors and D-factors of
  the NFW should prove particularly useful, as there are reasons
  increasing in quality and quantity~\citep[e.g.,][]{On15,Re15,Pe12,Pe10} for
  believing that many of the ultra-faints have pristine dark matter
  haloes, even if some of the largest dSphs most probably have cored
  haloes~\citep[e.g.,][]{Ag12,WP,Kl03,Am12}.}

We have computed the J-factors (and 1$\sigma$ uncertainties) for all
the known dSph galaxies with kinematic data. This includes the first
estimates for Horologium I and Grus I -- recently discovered in {\it
  Dark Energy Survey Data} data -- as well as Hydra II and Pisces II.
From this compendium, the ultrafaints with the highest J-factors are
Ursa Major II, Reticulum II, Tucana II and Horologium I.  They are
very attractive targets for indirect detection experiments. Although
our formulae assign Willman 1 and Segue 1 high J-factors, these
ambiguous objects may be extended clusters suffering disruption in the
Galactic tidal field rather than dwarf galaxies, perhaps complicating
or even invalidating previous estimates of $\sigmalos$ for these
systems.  Amongst the classical dSphs, Draco, Sculptor and Ursa Minor
have the highest J-factors.

If the J-factor profile can be mapped out as a function of integration
angle, then this raises the possibility of inferring properties for
the dark matter halo directly. We have shown for general halo models
that the J-factor rises from the origin, yet approaches a constant
asymptotic value with increasing integration angle. The behaviour at
the origin encodes information on the inner cusp slope $\gamma$. If
the model is cored or weakly cusped ($0\le \gamma \le 1/2$), then the
J-factor rises quadratically from the origin, $J\propto \theta^2$. If
the model is more strongly cusped ($1/2 < \gamma <3/2$), then $J
\propto \angle^{3-2\gamma}$. The asymptotic value of the J-factor --
which can be worked out analytically -- depends on both inner slope
$\gamma$, outer slope $\beta$ as well as the halo scale length and
normalisation.

Finally, we have identified a sweet spot, or preferential integration
angle, at which the J-factor is robust against changes in the dark
halo model. This is at the integration angle $\theta = \rh/D$, or the
ratio of the dSphs half-light radius $\rh$ to distance from the
observer $D$. This result holds good for models that exactly obey the
empirical relation between velocity dispersion and mass enclosed
within half-light radius~\cite{Wa09,Wo10}.

An important outstanding problem is the extension of this work to
flattened geometries. Most dSphs are flattened -- and some of the most
promising targets such as Reticulum II or Horologium I are very highly
flattened. Spherical models can provide useful guides, especially for
the largest classical dSphs (like Leo I or Fornax) that look nearly
round on the sky. They are least useful for the flattened
ultrafaints. In a companion paper, we show how to extend the
techniques presented here to explore flattened and triaxial
geometries~\citep{Sa16}.

\bigskip
\begin{acknowledgments}
{\red The authors thank the anonymous referees for helpful reports, as well
as Vasily Belokurov and Matt Walker for useful discussions. JLS
acknowledges financial support from the Science and Technology
Facilities Council (STFC) of the United Kingdom.
}

\end{acknowledgments}

\label{lastpage}


\begin{thebibliography}{99}

\bibitem[Kleyna et al.(2002)]{Kl02} Kleyna, J., Wilkinson,
M.~I., Evans, N.~W., Gilmore, G., \& Frayn, C.\ 2002, \mnras, 330, 792

\bibitem[Mateo(1998)]{Ma98} Mateo, M.~L.\ 1998, \araa, 36, 435

\bibitem[Colafrancesco et al.(2007)]{Co07} Colafrancesco,
S., Profumo, S., \& Ullio, P.\ 2007, \prd, 75, 023513

\bibitem[Culverhouse et al.(2006)]{Cu06} Culverhouse, T.~L.,
Evans, N.~W., \& Colafrancesco, S.\ 2006, \mnras, 368, 659

\bibitem[Evans et al.(2004)]{Ev04} Evans, N.~W., Ferrer, F.,
\& Sarkar, S.\ 2004, \prd, 69, 123501

\bibitem[Lake(1990)]{La90} Lake, G.\ 1990, \nat, 346, 39

\bibitem[Bergstr{\"o}m et al.(1998)]{Be98} Bergstr{\"o}m,
L., Ullio, P., \& Buckley, J.~H.\ 1998, Astroparticle Physics, 9, 137

\bibitem[Abdo et al.(2010)]{Ab10} Abdo, A.~A., Ackermann, M., Ajello,
  M., et al.\ 2010, \apj, 712, 147

\bibitem[Charbonnier et al.(2011)]{Ch11} Charbonnier, A.,
Combet, C., Daniel, M., et al.\ 2011, \mnras, 418, 1526

\bibitem[Walker et al.(2011)]{Wa11} Walker, M.~G., Combet,
C., Hinton, J.~A., Maurin, D., \& Wilkinson, M.~I.\ 2011, \apjl, 733, L46

\bibitem[Bonnivard et al.(2015a)]{Bo15a} Bonnivard, V.,
Combet, C., Maurin, D., \& Walker, M.~G.\ 2015a, \mnras, 446, 3002

\bibitem[Geringer-Sameth et al.(2015b)]{Ge15b} Geringer-Sameth, A.,
  Koushiappas, S.~M., \& Walker, M.\ 2015b, \apj, 801, 74

\bibitem[Martinez(2015)]{Mar15} Martinez, G.~D.\ 2015, \mnras, 451,
  2524

\bibitem[Zucker et al.(2006)]{Zu06} Zucker, D.~B.,
Belokurov, V., Evans, N.~W., et al.\ 2006, \apjl, 650, L41

\bibitem[Deason et al.(2012)]{De12} Deason, A.~J., Belokurov, V.,
  Evans, N.~W., Watkins, L.~L., \& Fellhauer, M.\ 2012, \mnras, 425,
  L101

\bibitem[Martin et al.(2008)]{Ma08} Martin, N.~F., de Jong, J.~T.~A.,
  \& Rix, H.-W.\ 2008, \apj, 684, 1075

\bibitem[Koposov et al.(2015)]{Ko15} Koposov, S.~E.,
Belokurov, V., Torrealba, G., \& Evans, N.~W.\ 2015, \apj, 805, 130

\bibitem[Laporte et al.(2013)]{La13} Laporte, C.~F.~P.,
Walker, M.~G., \& Pe{\~n}arrubia, J.\ 2013, \mnras, 433, L54

\bibitem[Sanders et al.(2016)]{Sa16} Sanders J.L., Evans N.W.,
  Geringer-Sameth A., Dehnen W.\ 2016, Phys Rev D, submitted.

\bibitem[O{\~n}orbe et al.(2015)]{On15} O{\~n}orbe, J.,
Boylan-Kolchin, M., Bullock, J.~S., et al.\ 2015, \mnras, 454, 2092

\bibitem[Read et al.(2015)]{Re15} Read, J.~I., Agertz, O.,
\& Collins, M.~L.~M.\ 2015, arXiv:1508.04143

\bibitem[Navarro et al.(1997)]{NFW} Navarro, J.~F., Frenk,
C.~S., \& White, S.~D.~M.\ 1997, \apj, 490, 493

\bibitem[Pe{\~n}arrubia et al.(2012)]{Pe12} Pe{\~n}arrubia,
J., Pontzen, A., Walker, M.~G., \& Koposov, S.~E.\ 2012, \apjl, 759, L42

\bibitem[Pe{\~n}arrubia et al.(2010)]{Pe10} Pe{\~n}arrubia,
J., Benson, A.~J., Walker, M.~G., et al.\ 2010, \mnras, 406, 1290

\bibitem[Agnello \& Evans(2012)]{Ag12} Agnello, A., \& Evans,
  N.~W.\ 2012, \apjl, 754, L39

\bibitem[Walker et al.(2009)]{Wa09} Walker, M.~G., Mateo,
M., Olszewski, E.~W., et al.\ 2009, \apj, 704, 1274

\bibitem[Wolf et al.(2010)]{Wo10} Wolf, J., Martinez, G.~D.,
Bullock, J.~S., et al.\ 2010, \mnras, 406, 1220

\bibitem[Walker
\& Pe{\~n}arrubia(2011)]{WP} Walker, M.~G., \& Pe{\~n}arrubia, J.\ 2011, \apj, 742, 20

\bibitem[Boyarsky et al.(2010)]{Bo10} Boyarsky, A.,
Ruchayskiy, O., Iakubovskyi, D., et al.\ 2010, \mnras, 407, 1188

\bibitem[Hernquist(1990)]{He90} Hernquist, L. 1990, \apj, 356, 359

\bibitem[Evans \& Williams(2014)]{Ev14} Evans, N.~W., \& Williams,
  A.~A.\ 2014, \mnras, 443, 791

\bibitem[McConnachie(2012)]{Mc12} McConnachie, A.~W.\ 2012,
\aj, 144, 4

\bibitem[Kirby et al.(2013)]{Ki13} Kirby, E.~N.,
Boylan-Kolchin, M., Cohen, J.~G., et al.\ 2013, \apj, 770, 16

\bibitem[Martin et al.(2015)]{Ma15} Martin, N.~F., Nidever, D.~L.,
  Besla, G., et al.\ 2015, \apjl, 804, L5

\bibitem[Belokurov et al.(2010)]{Be10} Belokurov, V.,
Walker, M.~G., Evans, N.~W., et al.\ 2010, \apjl, 712, L103

\bibitem[Kirby et al.(2015)]{Ki15} Kirby, E.~N., Simon,
J.~D., \& Cohen, J.~G.\ 2015, \apj, 810, 56

\bibitem[Walker et al.(2016)]{Wa15} Walker, M.~G., Mateo,
M., Olszewski, E.~W., et al.\ 2016, \apj, 819, 53

\bibitem[Amorisco \& Evans(2011)]{Am11} Amorisco, N.~C., \& Evans,
  N.~W.\ 2011, \mnras, 411, 2118

\bibitem[Bonnivard et al.(2015)]{Bo15c} Bonnivard, V., Combet, C.,
  Daniel, M., et al.\ 2015, \mnras, 453, 849

\bibitem[Ackermann et al. (2014)]{FermiLAT} Ackermann M., et al.,
  2014, PhRvD, 89, 042001

\bibitem[Koposov et al.(2011)]{Ko11} Koposov, S.~E., 
Gilmore, G., Walker, M.~G., et al.\ 2011, \apj, 736, 146 
  
\bibitem[Niederste-Ostholt et al.(2009)]{NO09} Niederste-Ostholt, M.,
  Belokurov, V., Evans, N.~W., et al.\ 2009, \mnras, 398, 1771

\bibitem[Bonnivard, Maurin, \& Walker (2015)]{Bo15b} Bonnivard V.,
  Maurin D., Walker M.~G., 2015, arXiv, arXiv:1506.08209

\bibitem[Willman et al. (2011)]{Willman1} Willman B., Geha M., Strader
  J., Strigari L.~E., Simon J.~D., Kirby E., Ho N., Warres A., 2011,
  \aj, 142, 128

\bibitem[Bechtol et al.(2015)]{Be15} Bechtol, K., Drlica-Wagner, A.,
  Balbinot, E., et al.\ 2015, \apj, 807, 50

\bibitem[Koposov et al.(2015)]{Ko15a} Koposov, S.~E., Casey,
A.~R., Belokurov, V., et al.\ 2015, \apj, 811, 62

\bibitem[Kleyna et al.(2003)]{Kl03} Kleyna, J.~T.,
Wilkinson, M.~I., Gilmore, G., \& Evans, N.~W.\ 2003, \apjl, 588, L21

\bibitem[Cole et al.(2012)]{Co12} Cole, D.~R., Dehnen, W.,
Read, J.~I., \& Wilkinson, M.~I.\ 2012, \mnras, 426, 601

\bibitem[Amorisco
\& Evans(2012)]{Am12} Amorisco, N.~C., \& Evans, N.~W.\ 2012, \mnras, 419, 184

\bibitem[Amorisco et al.(2013)]{Am13} Amorisco, N.~C.,
Agnello, A., \& Evans, N.~W.\ 2013, \mnras, 429, L89

\bibitem[Zhao(1996)]{Zh96} Zhao, H.\ 1996, \mnras, 278, 488

\bibitem[Han et al.(2016)]{Ha16} Han, J., Wang, W., Cole,
S., \& Frenk, C.~S.\ 2016, \mnras, 456, 1003

\end{thebibliography}
\end{document}